\documentclass[twocolumn,floatfix,amsmath,amssymb,letter,prb,citeautoscript,showpacs]{revtex4}

\usepackage{graphicx}
\usepackage{color}
\usepackage{amssymb}
\usepackage{amsmath}

\begin{document}

\author{Karthik Sasihithlu}%
\affiliation{%
Department of Mechanical Engineering, Columbia University\\
New York, NY 10027
}%
\author{Arvind Narayanaswamy}
\email{arvind.narayanaswamy@columbia.edu}
\affiliation{%
Department of Mechanical Engineering, Columbia University\\
New York, NY 10027
}%

\date{\today}



\title{Proximity Effects in Radiative Transfer}

\begin{abstract}
%
Though the dependence of near--field radiative transfer on the gap between two planar objects is well understood, that between curved objects is still unclear. We show, based on the analysis of the surface polariton mediated radiative transfer between two spheres of equal radii $R$ and minimum gap $d$, that the near--field radiative transfer scales as $R/d$ as $d/R \rightarrow 0$  and as  $\ln\left(R/d\right)$ for larger values of $d/R$ up to the far--field limit. We propose a modified form of the proximity approximation to predict near--field radiative transfer between curved objects from simulations of radiative transfer between planar surfaces.

\end{abstract}

\maketitle
For more than a century, Planck's theory of blackbody radiation and the radiative transfer theory (RTT) have been successful at predicting radiative heat transfer between objects when all length scales involved are much larger than the characteristic thermal wavelength $\lambda_T$ given by Wien's displacement law ($\lambda_T \approx 1.27 \hbar c/ k_B T$, where $2\pi\hbar$  is the Planck's constant, $c$ is the speed of light in vacuum, $k_B$ is the Boltzmann constant and $T$ is the absolute temperature). However, at shorter length scales, radiative transfer between two objects can exceed the predictions of Planck's theory due to electromagnetic near--field effects. Recent precision experiments \cite{narayanaswamy08a,shen09a,rousseau2009a} for measuring radiative heat transfer between a silica microsphere and a planar silica substrate have begun to shed new light on the enhancement of radiative transfer at nanoscale gaps due to surface phonon--polaritons. While near--field enhancement between parallel surfaces is well understood theoretically \cite{polder71,carminati99a,mulet02a,volokitin04a}, that between curved surfaces is still unclear.
Though near--field radiative exchange between a nanosphere and a plane has been calculated under different approximations\cite{mulet01a,carrillo2010}, rigorous numerical computation of near--field radiative transfer between a sphere of arbitrary diameter and a planar surface has not yet been possible due to computational difficulties. 
In order to understand the effect of curvature on enhancement of radiative transfer at nanoscale gaps, we investigate the near--field radiative transfer between two spheres of equal radii by rigorous simulations based on the dyadic Green's function technique \cite{tsang00a} and fluctuation--dissipation theorem \cite{rytov59a}.

 The problem of computing fluctuation--induced interaction between objects with curved surfaces is also encountered in literature on van der Waals and Casimir forces. A frequently used strategy to obtain forces between spherical surfaces, when the force between two parallel surfaces is known as a function of separation, is to use the so--called proximity approximation (or Derjaguin approximation) \cite{derjaguin1934untersuchungen}. While a rigorous proof for the proximity approximation for forces is still lacking \cite{sernelius2009test}, it's accuracy in predicting Casimir force has been well investigated theoretically \cite{gies06a, blocki77, genet2008casimir} and is expected to be most accurate when $R \gg d$, where $R$ is a characteristic radius of the objects involved, and $d$ is the minimum gap between them. Experimental measurements of Casimir force \cite{lamoreaux97a, mohideen98} have confirmed the applicability of the proximity approximation when $R \gg d$.

A similar technique \cite{rousseau2009a} has been used to compute radiative transfer. Measurements from the Chen group \cite{narayanaswamy08a,shen09a} between a silica microsphere and a silica substrate in the range 30 nm to 10 $\mu$m seemed to show better agreement with theoretical predictions of heat transfer between two spheres \cite{narayanaswamy07b} and did not agree with the proximity approximation. Rousseau et al. \cite{rousseau2009a}, based on their measurements between a silica microsphere and a silica substrate, concluded that near--field radiative transfer agreed with the proximity approximation in the range 30 nm to 2.5 $\mu$m. There are no experiments between two silica spheres reported in literature. Though the phenomena of van der Waals force (including Casimir force) and near--field radiative transfer are fluctuation--induced, there are important differences. Radiative transfer has contributions from the infra--red (IR) portion of the electromagnetic spectrum whereas forces have larger contributions from  the visible and higher frequencies. Dispersion forces obey a power law behavior and decay rapidly to zero as gap between the interacting bodies increases, while radiative transfer has a finite value due to propagating waves at large gaps too. Because of these differences, it is not clear whether the proximity approximation, as it is used to compute dispersion forces, can be used to predict near--field radiative heat transfer between spherical surfaces. 

\begin{figure}
\begin{center}
\includegraphics[width=8.5cm]{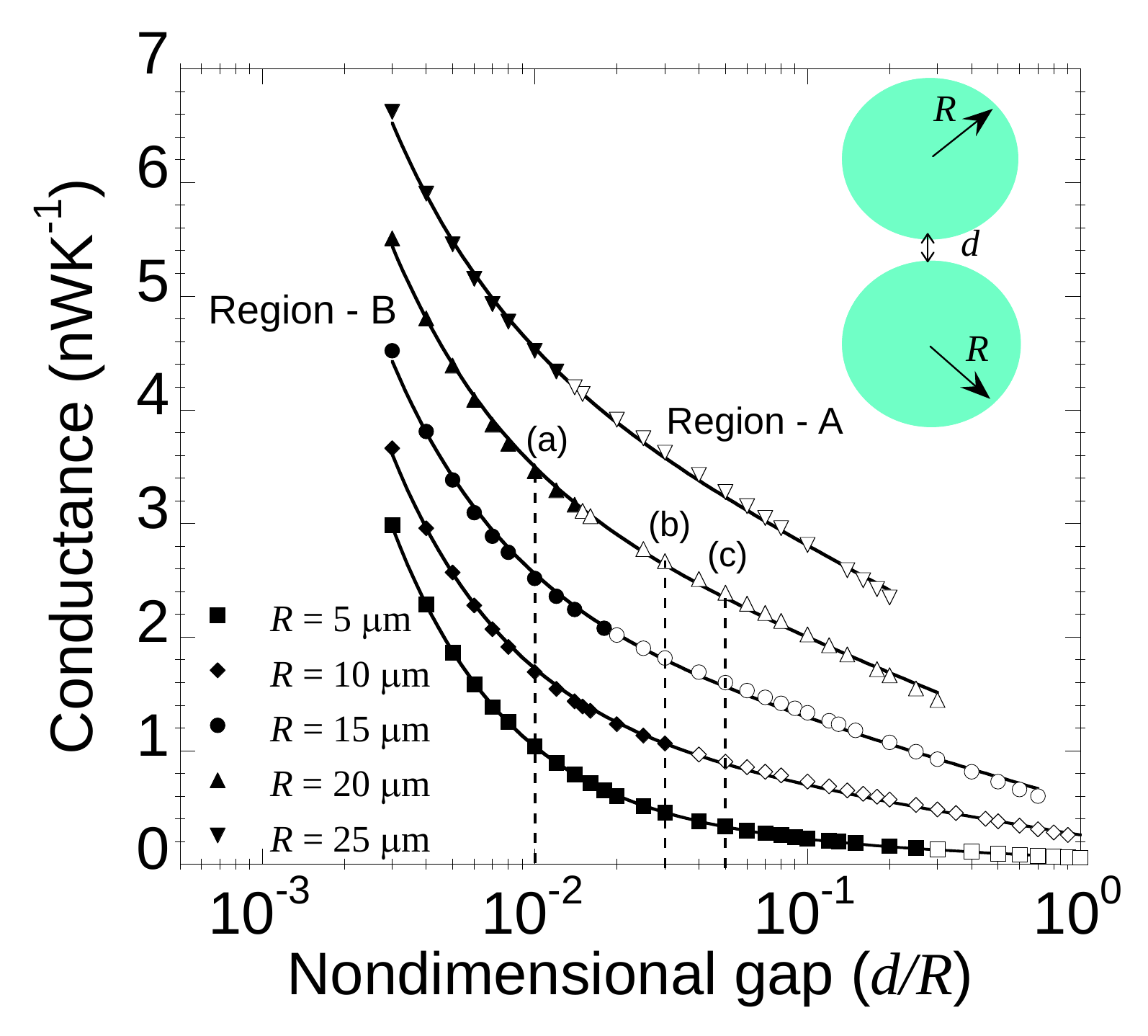}
\end{center}
\caption{\label{Fig1} Conductance between the two spheres shown in the top right corner as a function of $d/R$ for different radii. The open circles denote the conductance values which show a logarithmic variation with gap (marked Region--A) and the closed circles denote the conductance values which show a deviation from logarithmic behavior (marked Region--B). The spectral variation of the conductance at gaps marked (a), (b) and (c) for $R$ = 20 $\mu$m spheres are shown in \ref{Fig2}. }
\end{figure}

The configuration of the two spheres between which radiative transfer is to be calculated is shown in \ref{Fig1} (top right corner). Radiative heat transfer between the spheres is calculated using Rytov's theory of fluctuational electrodynamics \cite{rytov59a}. The Fourier component of the fluctuating electric field $\boldsymbol{E}(\boldsymbol{r}_{1}, \omega)$ and magnetic field $\boldsymbol{H}(\boldsymbol{r}_{1}, \omega)$ at any point $\boldsymbol{r}_{1}$ is given by \cite{tsang00a}:
\begin{eqnarray}
\label{eqn:EHGreen}
\boldsymbol{E}(\boldsymbol{r}_{1}, \omega) & = & i\omega \mu_{o} \int_{V} d^{3}r\overline{\overline{\boldsymbol{G}}}_{e}(\boldsymbol{r}_{1}, \boldsymbol{r}, \omega) \cdot \boldsymbol{J}(\boldsymbol{r}, \omega), \\
\boldsymbol{H}(\boldsymbol{r}_{1}, \omega) & = & \int_{V} d^{3}r \overline{\overline{\boldsymbol{G}}}_{h}(\boldsymbol{r}_{1}, \boldsymbol{r}, \omega) \cdot \boldsymbol{J}(\boldsymbol{r}, \omega),
\end{eqnarray}
where $\overline{\overline{\boldsymbol{G}}}_{e}(\boldsymbol{r}_{1}, \boldsymbol{r}, \omega)$ and $\overline{\overline{\boldsymbol{G}}}_{h}(\boldsymbol{r}_{1}, \boldsymbol{r}, \omega)$ are the dyadic Green's functions for the electric and magnetic fields due to a point source at $\boldsymbol{r}$ and are related by $\overline{\overline{\boldsymbol{G}}}_{h}(\boldsymbol{r}_{1}, \boldsymbol{r},\omega ) = \boldsymbol{\nabla} \times \overline{\overline{\boldsymbol{G}}}_{e}(\boldsymbol{r}_{1}, \boldsymbol{r}, \omega)$,  $\boldsymbol{J}(\boldsymbol{r}, \omega)$ is the Fourier component of the current density due to thermal fluctuations, and $\mu_{o}$ is the permeability of vacuum. The integration is performed over the entire volume $V$ containing the source.

The ensemble average Poynting vector $\boldsymbol{S} = \frac{1}{2}\operatorname{Re}\langle \boldsymbol{E} \times \boldsymbol{H}^{*}\rangle $ depends on the cross spectral densities of the components of the fluctuating current source, which are related to temperature by the fluctuation--dissipation theorem. Further details on the analysis can be found in Ref. 19. Refinement of the numerical method has enabled us to probe lower gaps and suggest better models for near--field radiative transfer. To investigate the effects of surface phonon--polaritons on near--field radiative transfer, silica has been the material of choice in experiments for two reasons: (1) it can support surface phonon--polaritons in the frequency ranges from 0.055 to 0.07 eV and 0.114 to 0.16 eV, and (2) silica microspheres are easily available. Hence the heat transfer has been computed, using silica as the material, for the frequency range 0.041 eV to 0.16 eV. The optical properties of silica are taken from Ref. 20. All numerical simulations have been conducted at 300 K.



\begin{figure}
\begin{center}
\includegraphics[width=8.5cm]{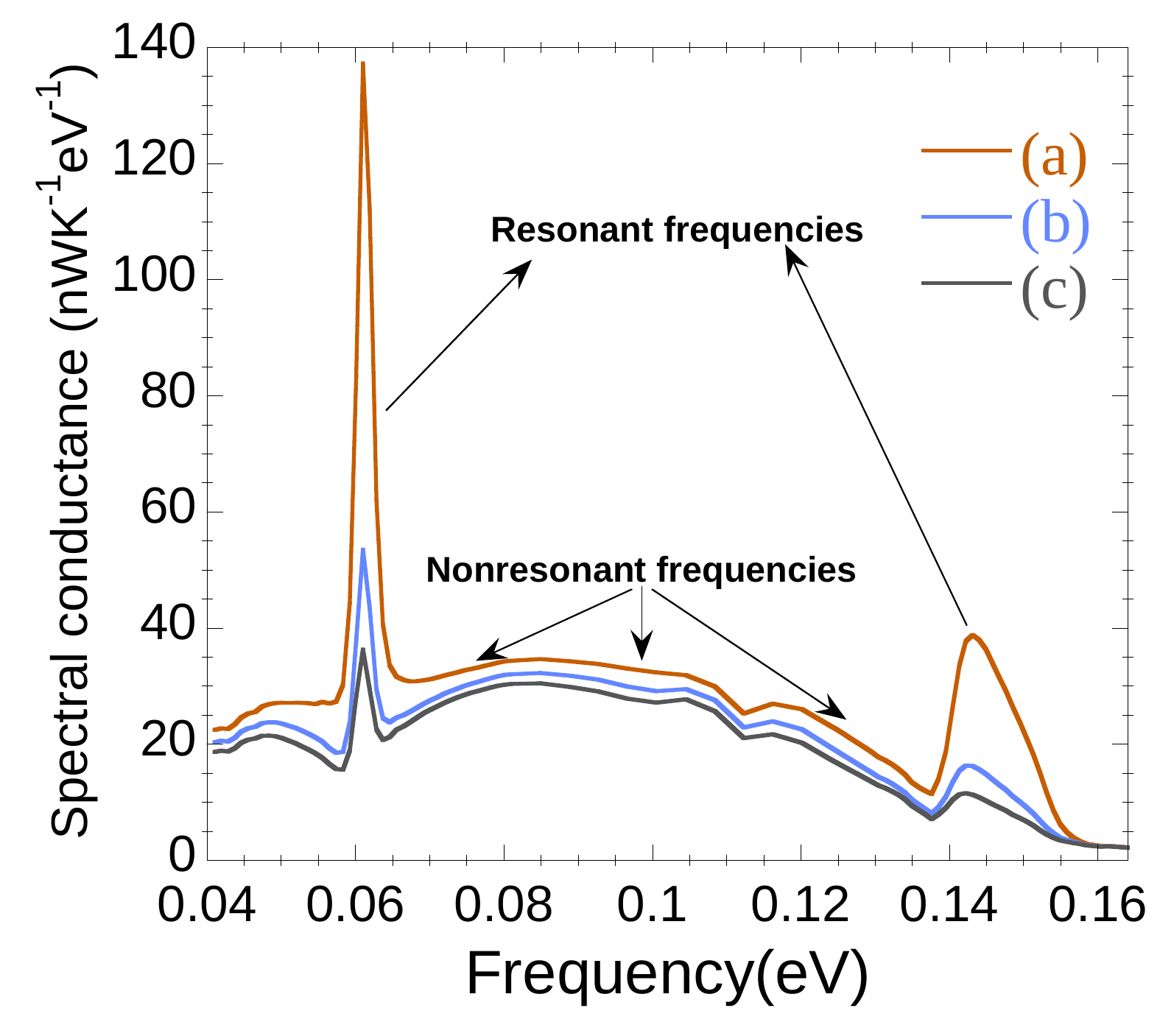}
\end{center}
\caption{\label{Fig2} The spectral variation of conductance for $R = 20\,\mu$m for the different gaps (a), (b) and (c) marked in \ref{Fig1}. The frequency regions marked ``Resonant frequencies'' (``Nonresonant frequencies'') are where surface phonon--polaritons are present (absent).}
\end{figure}

The linearized thermal conductance $G$ (WK$^{-1}$) between the two spheres is defined as:

\begin{equation}
\label{eq:conductance}
G =  \lim_{T_{1} \to T_{2}} \frac{Q (T_{1},T_{2})}{T_{1} - T_{2}},
\end{equation}
%
%
where $Q\left(T_{1},T_{2}\right)$ is the rate of heat transfer between the two spheres at temperatures $T_{1}$ and $T_{2}$. Numerical values of conductance are plotted as a function of gap to radius ratio $d/R$ for different radii in \ref{Fig1}. Only the points which deviate significantly from RTT have been included (see supplementary information, Section 1,  for further details). In \ref{Fig1}, for every radius, two regions can be observed: (1) a region where conductance varies logarithmically (marked Region--A), and (2) a region where a deviation from logarithmic behavior is observed (marked Region--B). 
To gain a deeper insight into this behavior, we compare the spectral variation of the conductance at different gaps in the two regions. The gaps chosen are 
$d/R$ = 0.01, 0.03 and 0.05 for $R =$ 20 $\mu$m (marked (a), (b) and (c) in \ref{Fig1}). The gaps corresponding to $d/R =$  0.03, 0.05 fall in Region--A of \ref{Fig1}, while $d/R = 0.01$ falls in Region--B. From the spectral variation plotted in \ref{Fig2}, we note that most of the increase in the heat transfer for $d/R$ = 0.01 is due to the contributions from surface phonon--polaritons alone. Hence we conclude that the contribution from surface phonon--polaritons to the conductance starts to become significant in Region--B.


\begin{figure}
\begin{center}
\includegraphics[width=8.5cm]{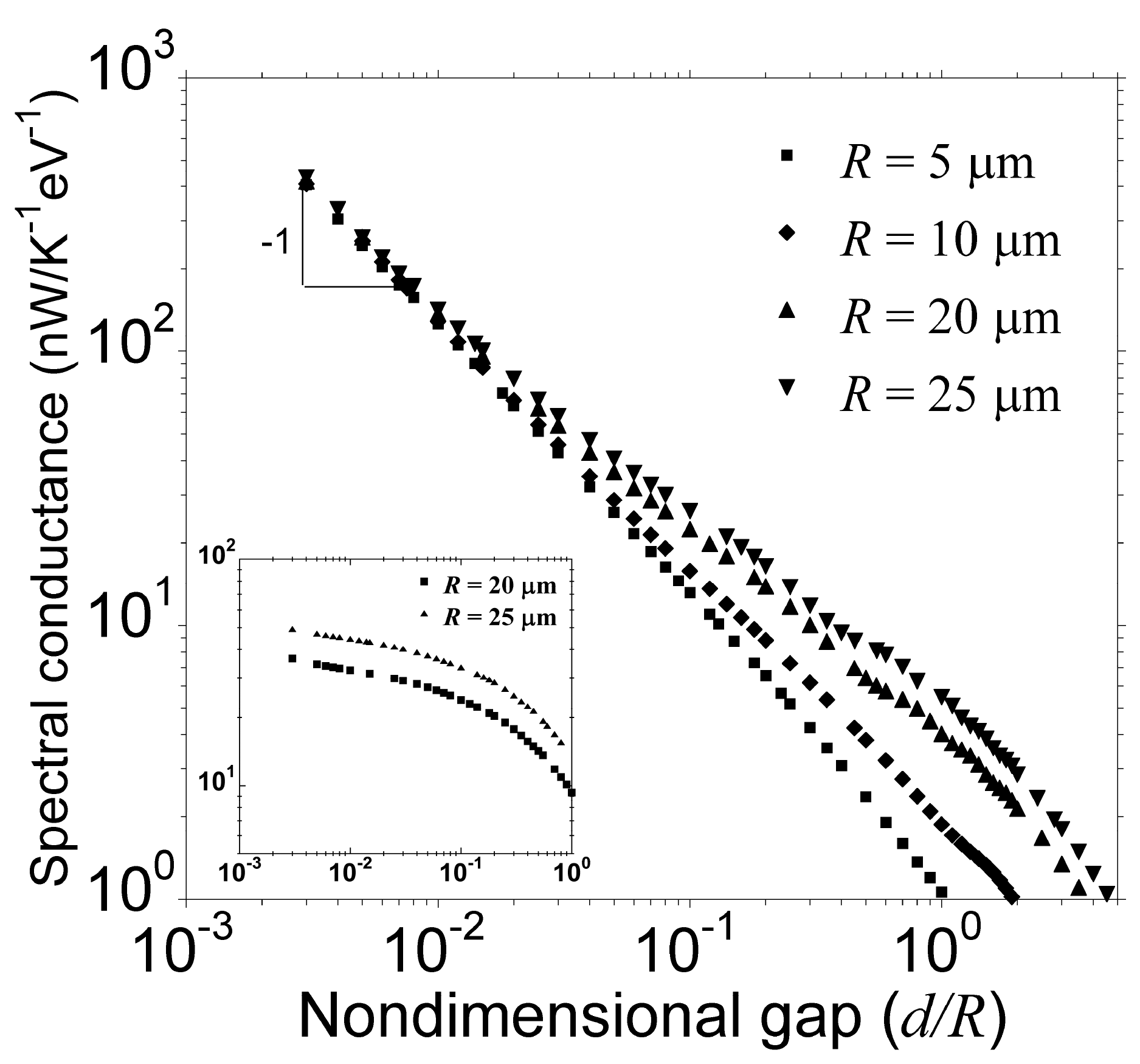}
\end{center}
\caption{\label{Fig3} For different radii, the spectral conductance at a resonant frequency (0.061 eV) as a function of $d/R$.  The conductance values for all radii attain a slope of $-1$ at low gaps. Inset: The spectral conductance at a nonresonant frequency (0.1005 eV) as a function of $d/R$ (axes labels remain the same).}
\end{figure}

We also analyzed the contributions of conductance from the resonant and nonresonant frequencies separately.  The results are shown in \ref{Fig3}. Remarkably, the analysis for the spectral conductance at a resonant frequency (0.061 eV) suggests that at gaps $d/R \lesssim 0.01$, the conductance is dependent only on the ratio $d/R$ and is independent of the particular values of $d$ and $R$. Furthermore, the slope of the data points being $\approx -1$ suggests a $\left(R/d\right)$ behavior at such gaps. A similar analysis for a nonresonant frequency (0.1005 eV), shown in the inset of \ref{Fig3}, suggests that for $d/R \lesssim 0.01$ the rate of change of spectral conductance with gap is significantly lower than that for resonant frequencies. 
Based on the behavior of resonant radiative transfer at small gaps and the observation of logarithmic behavior for larger gaps in  \ref{Fig1}, we have found that the numerical values of conductance can be modeled by a function of the form $C_{1}\left(R/d\right) + C_{2}\ln\left(R/d\right) + C_{3}$, where $C_{1}$, $C_{2}$ and $C_{3}$ are radius--dependent constants. $C_{1}$, $C_{2}$ and $C_{3}$ are obtained by minimizing the square of error between the function and the numerical values of conductance. The fitted curves are shown in  \ref{Fig1}. (For details on the variation of $C_{1}$, $C_{2}$ and $C_{3}$ with $R$ see supplementary material, Fig. 2). 

\begin{figure}
\begin{center}$
\begin{array}{cc} 
\includegraphics[width=5cm]{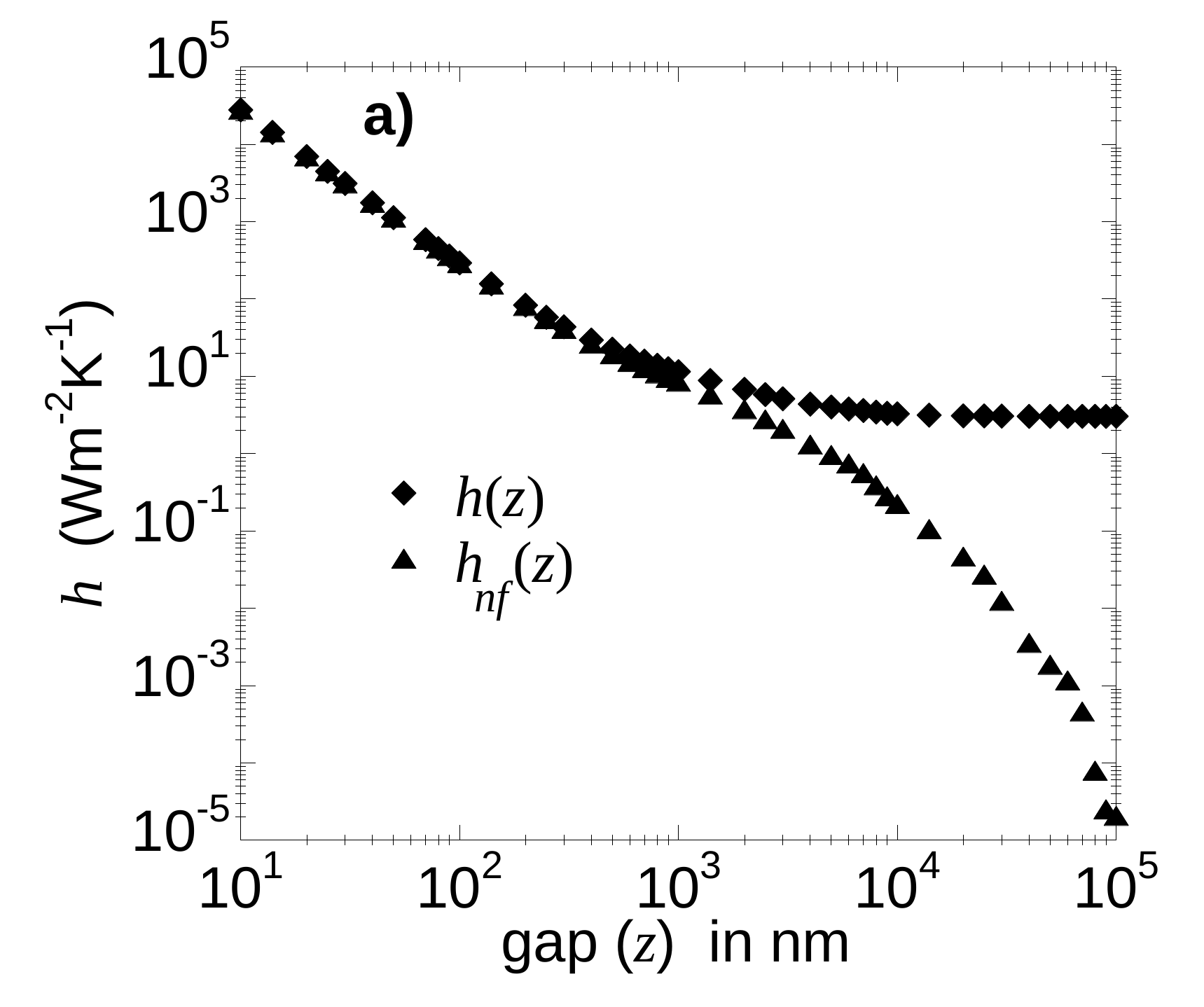} &
\includegraphics[width=3.5cm]{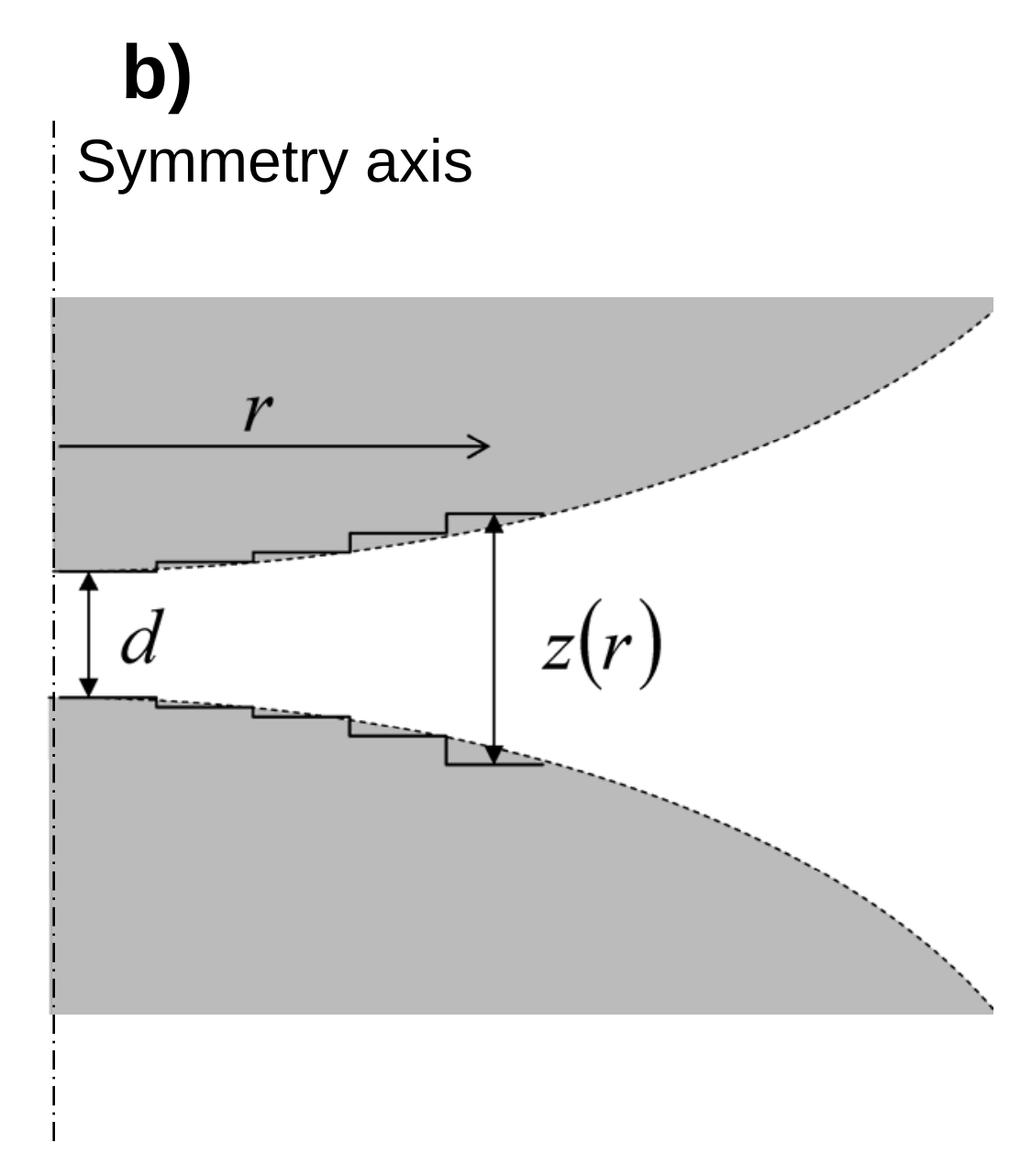}
\end{array}$
\includegraphics[width=8.5cm]{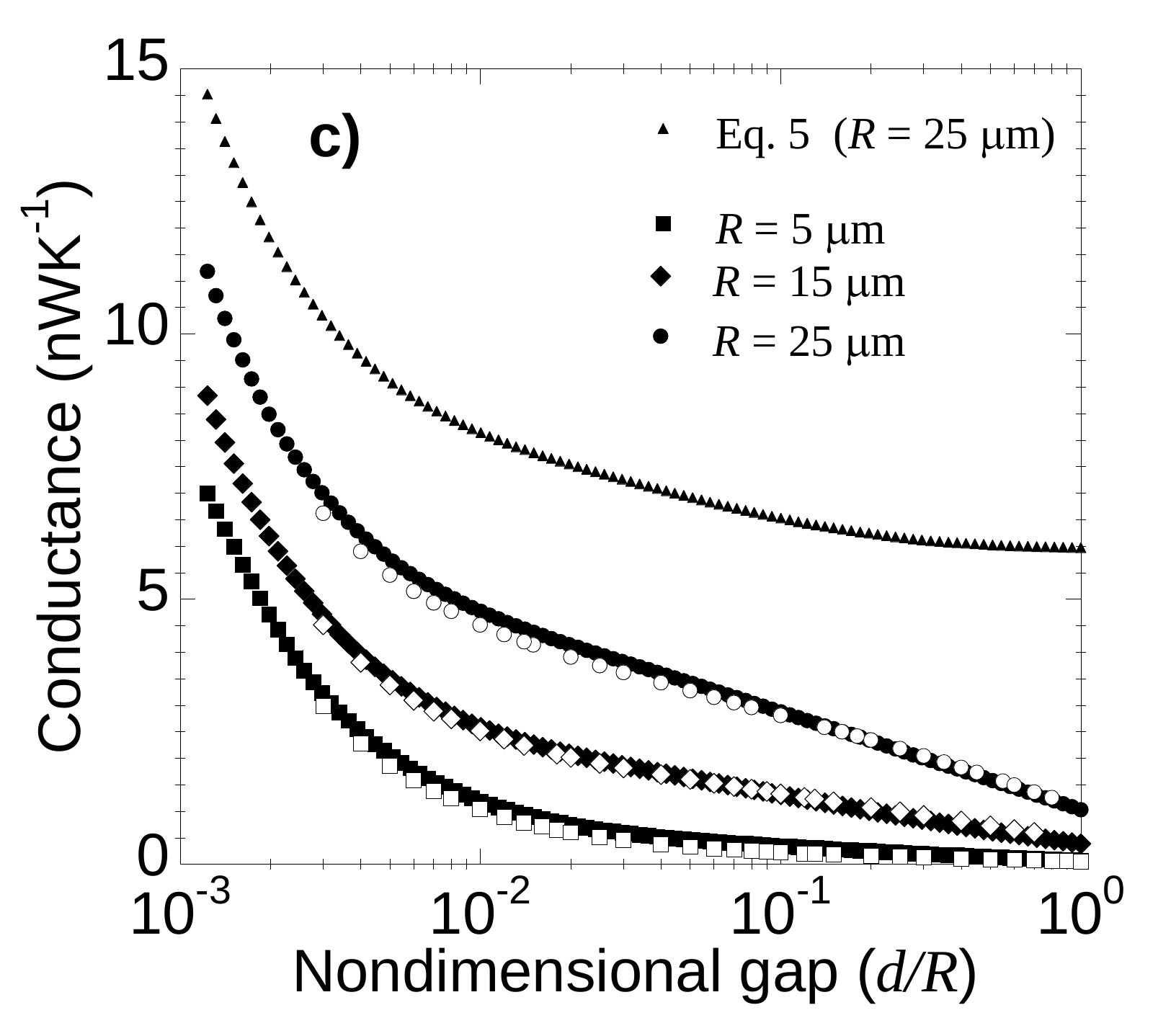}
\end{center}
\caption{\label{Fig4} {\bf (a)} Variation of $h$($z$) and $h_{nf}$($z$) with gap between two flat silica surfaces. {\bf (b)} Proximity approximation -- the conductance between two spheres is calculated by summing the local contributions of the heat transfer coefficient between two parallel planes. {\bf (c)} Comparison between the conductance values obtained numerically and using proximity approximation. The open cirlces denote the numerical values while the closed circles denote the proximity approximation predictions using  \ref{eq:prox}. Proximity approximation using  \ref{eq:Greffet} for $R$ = 25 $\mu$m has been included for comparison. }
\end{figure}

As has been mentioned earlier, conductance between curved surfaces can also be estimated from the known solutions for near--field radiative transfer between parallel surfaces using the proximity approximation. The heat transfer coefficient $h(z)$ for two flat silica surfaces is plotted as a function of gap $z$ in  \ref{Fig4}(a). 
$h(z)$ can be split as follows:

\begin{equation}
\label{eqn:splitparallelHTC}
h\left(z\right) = h_{nf}\left(z\right) + h_{\infty},
\end{equation}
where $h_{\infty}$ is the contribution from propagating waves, which  attains a constant value for $z \gg \lambda_{T}$, and $h_{nf}$ contains contributions from all other effects. As can be seen from  \ref{Fig4}(a), $h_{nf}\left(z\right) \rightarrow 0$ as $z \rightarrow \infty$.

Rousseau et al. \cite{rousseau2009a} applied the proximity approximation to compute the theoretical conductance between a sphere and a flat surface. Applying the same form of proximity approximation for finding the conductance between two spheres, we get:  

\begin{equation}
\label{eq:Greffet}
G(d,T)=\int_0^R\!h(z) 2 \pi r \, dr,
\end{equation}
where $z = d + 2R - 2 \sqrt{R^2 - r^2}$ is the local gap between the two spheres as shown in  \ref{Fig4}(b) and $h(z)$ is the heat transfer coefficient between two parallel surfaces at that gap.  From  \ref{Fig4}(a), we observe that $h_{nf} >> h_\infty$ for gaps $z \lesssim 400$ nm. 
When sizes of spheres are such that $z(r) \lesssim 400$ nm for $0 \le r \le R$, most of the contribution to the conductance is from near--field effects. However, for sizes of spheres currently used in experiments ($ R > 5$ $\mu$m) there will be significant contribution from larger gaps where near--field effects are minimal.  
While \ref{eq:Greffet} might provide a reasonable approximation to the contribution from near--field effects to the radiative transfer, it overestimates the contribution from propagating waves. In the large--gap limit, where the  heat transfer  coefficient attains a constant value, $h_\infty$,  \ref{eq:Greffet} predicts conductance between the spheres to be $\pi R^2 h_\infty$, irrespective of the gap and does not take into account the variation of view factor \cite{siegel2002thermal} between the two spheres with distance.

In lieu of this we propose that the proximity approximation, in the form that is used to determine Casimir or van der Waals forces between spherical surfaces, be used to predict only the contribution to conductance from $h_{nf}$. The contribution to the conductance from propagating waves is computed according to RTT by taking into account the variation of view factor with distance between the two spheres. This correction to the proximity approximation formulation is not necessary while calculating Casimir or van der Waals force, since they decay rapidly with distance ($1/d^4$ and $1/d^3$ respectively). The modified form of proximity approximation to determine the conductance is:

\begin{equation}
\label{eq:prox}
G\left(d,T\right)=\int_0^R\!h_{nf}(z) 2\pi r \, dr\,+\, G_{c}(d,T),
\end{equation}
where $G_{c}(d,T)$ can be approximated by the conductance value from RTT when diffraction effects are negligible. $G_{c}(d,T)$ for two objects of equal emissivity $\epsilon$ and surface area $A$ is given by \cite{siegel2002thermal}: 

\begin{equation}
 \label{eq:supp_G} G_{c}(d,T) =  \frac{4 \sigma A T^{3}}{2 (1 - \epsilon)/\epsilon + (1/F_{12})},
\end{equation}
where $F_{12}$ is the view factor between the two objects. 
Conductance values computed using  \ref{eq:prox} and  \ref{eq:supp_G} are in greater agreement with the numerically computed values than the prediction of  \ref{eq:Greffet} as shown in  \ref{Fig4}(c). Since the conductance values have been computed for the frequency range 0.041 eV to 0.164 eV,  $G_{c}(d,T)$  has been determined for this frequency range too (see supplementary material, Section 1, for more details). Numerical values of the gap dependent view factor between the two spheres is taken from Ref. 22. 

 While the modified form of the proximity approximation can also be used to predict the conductance between a sphere and a flat surface, there are no rigorous numerical simulations of near--field radiative transfer between a sphere and a flat surface to compare them with. If this can be accomplished, the validity of this modified proximity approximation for predicting the conductance between sphere and a plane can be verified. Another configuration that can be used to test the predictive capability of the modified proximity approximation is that between two parallel cylinders. Current experimental setups are not ideal for comparison either. Since the sphere is positioned close to the edge of the substrate due to experimental constraints, deviation from RTT could occur due to surface phonon--polariton mediated near--field effects as well as diffraction effects which need to be computed for each geometry separately.


In summary, radiative heat transfer between two spheres has been analyzed in the near--field regime using flucutational electrodynamics formalism. We have shown that it varies as $R/d$ as $d/R \rightarrow 0$ and as $\ln\left(R/d\right)$ for larger values of $d/R$ up to the far--field limit.
We have also shown that the proximity theorem, in the form that is used to compute dispersive forces cannot be used to determine near--field heat transfer and a modification is needed to take into account the contributions from propagating waves. Further numerical simulations of near--field radiative transfer between curved surfaces are necessary to test the predictive capability of the modified proximity approximation in more general configurations than the one considered here. Such predictive capabilities will be useful in applications like heat--assisted magnetic recording \cite{challener2009a} to approximate the near--field radiative transfer between the curved surface of the near--field transducer and the recording medium. 


This work was financially supported by NSF (Grant No. NSF CBET--0853723).

\providecommand*\mcitethebibliography{\thebibliography}
\csname @ifundefined\endcsname{endmcitethebibliography}
  {\let\endmcitethebibliography\endthebibliography}{}

\end{document}